\documentclass[aps,pra,twocolumn,showpacs,groupedaddress]{revtex4-1}

\usepackage{amsfonts,mathrsfs,amsmath,amsthm,amssymb,bbold}
\usepackage{epsfig}
\usepackage{bm}
\usepackage{tabulary}
\usepackage{color}
\usepackage[dvipsnames]{xcolor}
\usepackage{hyperref}

\begin{document}  
\title {\bf Improving direct state measurements by using rebits in real enlarged Hilbert spaces }

\author{ Le Bin Ho}
\thanks{Electronic address: binho@kindai.ac.jp}

\thanks{\\$\copyright$2018. This manuscript version is made available under the CC-BY-NC-ND 4.0 license \href{http://creativecommons.org/licenses/by-nc-nd/4.0/}{http://creativecommons.org/licenses/by-nc-nd/4.0/}}
\affiliation{Department of Physics, Kindai University, Higashi-Osaka, 577-8502, Japan}
\affiliation{Ho Chi Minh City Institute of Physics, VAST, Ho Chi Minh City, Vietnam}

\date{\today}

\begin{abstract}
We propose a protocol to improve the accuracy of direct complex state measurements (DSM)
by using rebits in real Hilbert spaces. 
We show that to improve the accuracy, 
the initial complex state should be decomposed into the real and imaginary parts 
and stored in an extended state (rebit) which can be tracked individually by two bases of an extra qubit. 
For pure states, the numerical calculations show that 
the trace distances between the true state and the reconstructed state 
obtained from the rebit method are more precise than those ones obtained from 
the usual DSM and quantum state tomography (SQT)
because the number of projective measurements is reduced. 
For mixed states, the rebit method gives the same accuracy  
in comparison to the usual DSM, 
while it is less precise than QST.
Its precision is also significantly improved when using nearly-pure states.
Our proposal holds promises as a reliable tool for quantum computation, 
testing of quantum circuits by using only real amplitudes.
\end{abstract}

\maketitle

{\section{Introduction}\label{sec_i}}
A complex wave function in quantum mechanics plays a key role 
to understand natural phenomena at the quantum scale. 
It is the fundamental representation of the quantum state of a system. 
According to its statistical interpretation, the wave function enables one to 
predict the results of measurements made on the system. 
Therefore, the complete determination of the wave function, or in general, 
the density matrix, of quantum states is crucially important and is one of 
the main tasks in quantum mechanics. 

A practical technique for quantum state determination 
known as quantum state tomography (QST) has been proposed \cite{James64}. 
In this method, multiple copies of the system are measured 
in a complete set of noncommuting observables, 
thus it is said to be indirect. 
From the measurement results, the quantum state is reconstructed. 
QST recently has been improved by using  mutually unbiased bases (MUB)
\cite{Wootters191,Adam105,Durt8} and also has been experimentally demonstrated  
in high-dimensional quantum information \cite{Bent5}. 
It is, however, very difficult to apply for high-dimensional systems 
because it requires large calculation cost and very precise
measurements.

Another approach, which is named as direct state measurement (DSM) 
has been originally proposed by Lundeen {\it et al.} \cite{Lundeen474,Lundeen108} 
and has been extensively studied recently \cite{Bamer112,Sal7,Mir113, Malik5,Lorenzo110,Lorenzo88,Wu3,Bol7,Shi2,Shi,Thekkadath117,Chen381,Zhu376}. 
This method is based on the fact that 
the amplitudes of the wave functions are proportional to the weak values \cite{Lundeen474}.
DSM has more experimental merit than QST because it is straightforward, simple, versatile,
and also able to apply for large systems \cite{Shi2,Shi}.
 Furthermore, DSM only requires local measurements
 \cite{Lundeen108}. 
The accuracy of DSM based on weak measurement does not high 
in comparison to QST because of the bias errors caused 
by the weakness of the interactions during the measurements \cite{Maccone89}. 
However, Vallone {\it et al.} have pointed out that DSM can also be based on 
strong measurements and can offer more precise results \cite{Vallone116}.
The authors considered a measurement scheme with an arbitrary strength 
and analyzed the precision and accuracy of the weak DSM 
and strong DSM, which correspond to the weak and strong coupling strengths, respectively. 
They concluded that a better estimation of the wave function could be achieved 
by using strong measurements and therefore the weak measurements are not beneficial. 
Their proposal, recently, has been experimentally demonstrated in a matter-wave interferometer \cite{Denkmayr118}.  
 
In this work, we propose a protocol to improve the DSM scheme 
by using the ``rebit" in the real Hilbert space.
The prefix ``re" means ``real" in the Hilbert space. 
We name our method as ReDSM for short. 
In our proposal, we first decompose a complex wave function (of a pure state) 
or a density matrix (of a mixed state) into the real and imaginary parts 
and map to a rebit state by adding an extra qubit.
The enlarged system which is formed by the system and the extra qubit 
is coupled to a qubit pointer via an arbitrary coupling strength. 
After the interaction, the system is postselected onto the conjugate basis.
The projective measurements will be implemented in the two-qubit state outcome 
(the extra qubit and the pointer qubit) to reconstruct the quantum state.

The trace distance between the true state and the reconstructed state 
is used as a figure of merit of the accuracy. 
We first consider pure states and then
move to mixed states.
For mixed states, in the ReDSM case, we perform the projective measurements 
in both the pairwise combinations of eigenstates of the Pauli matrices 
which is referred as the standard separable quantum state basis (SSB ReDSM) 
\cite{Adam105} and  the Bell and Bell-like quantum states basis (BBB ReDSM).
We  compare the results with the usual DSM and the mutually unbiased bases 
quantum state tomography (MUB QST) \cite{Wootters191,Adam105,Durt8}. 
We then also consider the reconstruction of nearly-pure states.

The structure of this paper is organized as follows.  
Section \ref{sec_ii} introduces rebits and the evolution operator that operates on the rebits. 
Section \ref{sec_iii} presents the main procedure of our proposal  
and the numerical simulation results for pure states, mixed states, and also nearly-pure states.  
The paper concludes with a brief summary in Sec. \ref{sec_iv}. 

{\section{Rebit and its evolution}\label{sec_ii}}
It is well known that the universal quantum computing can be transformed to circuits that use only the real amplitudes \cite{Rudo,Bern}, e.g., the Quantum Turing Machine (QTM) model \cite{Adleman26}. The corresponding state that used in such transformation is a superposition of rebits, the {\it real versions} of qubits.  Recently, the content of the rebits has been widely used \cite{Delfosse5,Matthew102}. A rebit can be established as we describe below.  Given a quantum state in $d$-dimensional Hilbert space in the computational basis $\{|n\rangle\}$ as
\begin{align}\label{psi_n}
|\psi\rangle = \sum_{n=0}^{d-1}(\psi^r_n+i\psi^i_n)|n\rangle, 
\end{align}
where
\begin{align}\label{re-im}
\psi^r_n \equiv \langle n|{\rm Re}|\psi\rangle, \hspace{0.1cm} {\rm and} \hspace{0.3cm} \psi^i_n \equiv \langle n|{\rm Im}|\psi\rangle\;,
\end{align}
where ${\rm Re}|\psi\rangle$ and ${\rm Im}|\psi\rangle$ are real numbers. 
This is the standard form of the state. The real and imaginary parts of the state can be decomposed and stored in the two dimensions of an extra qubit as \cite{Matthew102}
\begin{align}\label{psi_til}
|\widetilde\psi\rangle = \sum_{n=0}^{d-1}\psi^r_n|n\rangle|0\rangle+\psi^i_n|n\rangle|1\rangle =
\begin{pmatrix}
\psi^r_1\\
\psi^i_1\\
\psi^r_2\\
\psi^i_2\\
\vdots
\end{pmatrix}\;,
\end{align}
which is known as the superposition state of the rebit or the enlarged state because $|\tilde \psi\rangle$ lives in 2$d$-dimension.
%
%
%
It is worthy to note that the state $|\widetilde\psi\rangle$ is the enlarged quantum state in embedding quantum simulators \cite{Candia111}. Such rebit state $|\widetilde\psi\rangle$ can be prepared initially in enlarged Hilbert spaces \cite{Candia111} and can be experimentally implemented in a photonics system \cite{Loredo116, Chen116}. 

For the evolution, each complex gate $\bm{U}$ operating on the state $|\psi\rangle$ will be replaced by its real form, denoted as $\bm{\widetilde U}$, operating on $|\widetilde\psi\rangle$.  $\bm{\widetilde U}$ is defined by \cite{Aha}
\begin{align}
\bm{\widetilde U} |n\rangle|0\rangle &= [{\rm Re}(\bm{U})|n\rangle]|0\rangle+[{\rm Im}(\bm{U})|n\rangle]|1\rangle,\label{u_til0}\\
\bm{\widetilde U} |n\rangle|1\rangle &= -[{\rm Im}(\bm{U})|n\rangle]|0\rangle+[{\rm Re}(\bm{U})|n\rangle]|1\rangle.\label{u_til1}
\end{align}
Here ${\rm Re}(\bm{U})$ and ${\rm Im}(\bm{U})$ are the real and imaginary parts of the operator $\bm{U}$, respectively. 
If $\bm{U}$ causes the evolution
\begin{align}\label{psi_bar}
|\overline\psi\rangle = \bm{U}|\psi\rangle = \sum_{n=0}^{d-1}(\bar\psi^r_n+i\bar\psi^i_n)|n\rangle,
\end{align}
then $\bm{\widetilde U}$ should operate the evolution as
\begin{align}\label{psi_bar_til}
|\overline{\widetilde\psi}\rangle = \bm{\widetilde U}|\widetilde\psi\rangle = \sum_{n=0}^{d-1}\bar\psi^r_n|n\rangle|0\rangle+\bar\psi^i_n|n\rangle|1\rangle\;.
\end{align}
Rudolph and Grover showed that the quantum gate $\bm{\widetilde U}$ satisfies the universal quantum computation \cite{Rudo}. Hereafter, let us give an example to illustrate that the evolution (\ref{psi_bar_til}) can be achieved by using the definitions (\ref{u_til0}, \ref{u_til1}). Consider an arbitrary single-qubit rotation in the $z$-direction, the gate is given as
\begin{align}\label{psi'}
R_z(\tau) = 
 \begin{pmatrix}
 1& 0 \\
 0& e^{i\tau} 
 \end{pmatrix}.
 \end{align}
The standard form \eqref{psi_n} and its evolved state explicitly yield
\begin{align}\label{psi_psi_bar}
\notag&|\psi\rangle = \psi_0^r|0\rangle +  i\psi_0^i|0\rangle +  \psi_1^r|1\rangle +  i\psi_1^i|1\rangle\\
\notag &\to \psi_0^r|0\rangle +  i\psi_0^i|0\rangle \\
 &+  (\cos\tau\psi_1^r-\sin\tau\psi_1^i)|1\rangle +  i(\sin\tau\psi_1^r+\cos\tau\psi_1^i)|1\rangle.
\end{align}
Four terms after the arrow correspond to $\bar\psi_0^r, \bar\psi_0^i, \bar\psi_1^r,$ and $\bar\psi_1^i$ in Eq. \eqref{psi_bar}, respectively. Meanwhile, the equivalent evolution of the enlarged form \eqref{psi_til} yields
\begin{align}\label{til_psi_psi_bar}
\notag&|\widetilde\psi\rangle = \psi_0^r|00\rangle +  i\psi_0^i|01\rangle +  \psi_1^r|10\rangle +  i\psi_1^i|11\rangle\\
\notag& \to \psi_0^r|00\rangle +  i\psi_0^i|01\rangle +  (\cos\tau\psi_1^r-\sin\tau\psi_1^i)|10\rangle \\
&+  i(\sin\tau\psi_1^r+\cos\tau\psi_1^i)|11\rangle,
\end{align}
where we have used the evolutions in Eqs. (\ref{u_til0}, \ref{u_til1}). 
The role of this enlarged evolution somehow plays the same role as quantum simulators \cite{Feynman21,Georgescu86,Buluta326}. 
In this context, one can replace the evolution of a quantum system 
[e.g. Eq. \eqref{psi_psi_bar}] by an alternative evolution which can be controlled 
and of course, satisfies the nature of the evolution of the quantum system 
[e.g., Eq. \eqref{til_psi_psi_bar}.]

Notable that our rebit method is typically different from 
a previous proposal of Ref. \cite{Allahverdyan92}. 
In Ref. \cite{Allahverdyan92}, a system $S$ interacts with 
an auxiliary system $A$ to determine the unknown state of $S$ 
using a single apparatus repeated measurements in the joint system $S+A$. 
While $S$ no need to be postselected in that method, 
it is required postselection in our rebit method.
Furthermore, the dimension of $A$ must be larger 
or equal to the dimension of $S$, 
which is usually wasted the resources and difficult to prepare in practice, 
whereas the extra system in the rebit method is a qubit (two dimensions.) 
The accuracy of the reconstruction in the previous method appears not well established and
depends on the interaction between $S$ and $A$ and the interaction time.
In contrast, as we will see later, the rebit method improves the accuracy 
of the reconstruction for pure states and nearly-pure states. 
For these reasons, we propose to use rebits for the quantum state tomography as following. 

{\section{Direct state measurement with rebits}\label{sec_iii}}
In this section, we use rebits to reconstruct complex quantum states for both cases pure and mixed states including nearly-pure states. 

\begin{figure} [t]
\centering
\includegraphics[width=8.6cm]{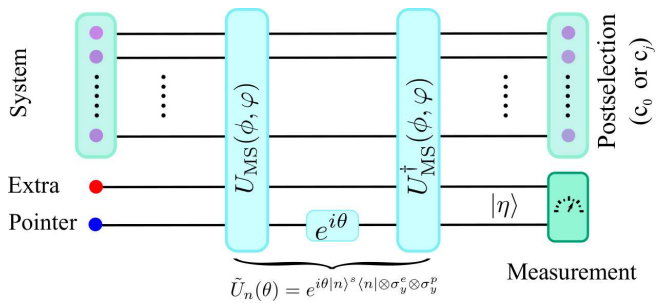}
\caption{
(Color online.) Quantum circuit implements the interaction scheme. An extra qubit is added to the system to track the real and imaginary parts of the system. This enlarged system then interacts with a qubit pointer. After the interaction, the system is postselected onto $|c_0\rangle$ for pure states or $|c_j\rangle$ for mixed states. The measurements of the entangled state $|\eta\rangle$ formed by the extra qubit and pointer qubit give the desire reconstructed state. The evolution $\bm{\widetilde U}_n(\theta)$ can be simulated by using the nonlocal entangling M\o lmer-S\o rensen gates and a local single-qubit rotation in the pointer (see Appendix \ref{appA}.)
}
\label{fig1}
\end{figure} 

{\subsection{For pure states}}
The measurement process can be depicted in a quantum circuit 
as shown in Fig. \ref{fig1}. 
Suppose that a quantum system is prepared in the initial state 
$|\psi\rangle$ as in Eq. \eqref{psi_n}. 
The system state is embedded onto an enlarged state by adding one extra qubit.
The enlarged state now becomes the rebit $|\tilde\psi\rangle$ as in Eq. \eqref{psi_til}. 
Notable that this rebit can be prepared even the initial system state is unknown 
 \cite{Candia111,Alvarez111}. 
The rebit $|\widetilde\psi\rangle$ is coupled to a pointer qubit 
which is prepared in the state $|0\rangle_p$, 
where the subscript $p$ stands  for ``pointer." 
The total state, therefore is given by 
$|\Psi_{\rm in}\rangle=|\widetilde\psi\rangle \otimes|0\rangle_p$. 
The total system evolves according to the following interaction operator \cite{Ho97} 
%
\begin{align}{\label{operator}}
\notag \bm{\widetilde U}_n(\theta)&=e^{i\theta|n\rangle\langle n|_s\otimes\sigma^y_e\otimes\sigma^y_p}\\
\notag&=\bm{I}_s\otimes \bm{I}_e\otimes \bm{I}_p + (\cos\theta-1)|n\rangle\langle n|_s \otimes \bm{I}_e\otimes \bm{I}_p \\
&+ i\sin\theta |n\rangle\langle n|_s\otimes\sigma^y_e\otimes\sigma^y_p,
\end{align}
%
where $\theta=gt$ stands for the coupling strength with $0<\theta\le\frac{\pi}{2}$. 
We also assume $\hbar=1$. 
The superscripts $s, e$ and $p$ mean ``system," ``extra," and ``pointer," respectively. 
See Appendix \ref{appA} for detailed implementation of $\bm{\widetilde U}_n(\theta)$.
After the interaction, the system is postselected onto the conjugate basis as 
$|c_0\rangle = 1/\sqrt{d}\sum_{n=0}^{d-1}|n\rangle$.
 The outcome, which is an entangled state between the extra qubit and pointer qubit, becomes 
\begin{align}{\label{final}}
\notag|\eta\rangle = \dfrac{1}{\sqrt{d}}\Bigl[&(\Sigma^r-\epsilon\psi^r_n)|00\rangle + (\Sigma^i-\epsilon\psi^i_n)|10\rangle\\
&-i\sin\theta\psi^r_n |11\rangle +i\sin\theta\psi^i_n |01\rangle \Bigr]_{e,p},
\end{align}
where $\Sigma^j \equiv \sum_{n=0}^{d-1}\psi_n^j$, ($j = r, i$), 
$\epsilon \equiv 2\sin^2\frac{\theta}{2}$. 
Obviously, by calculating the outcomes (1 or 0) of the extra qubit 
conditioned on the pointer outcome 1, 
we can completely obtain the real and imaginary parts 
of the given quantum state in the basis $\{|n\rangle\}$. 
In other words, the real and imaginary parts of the quantum state can be derived directly via
\begin{align}\label{state}
\psi^r_n =  \dfrac{\sqrt{dP_{11}}}{\sin\theta}\;, \hspace{0.1cm}{\rm and}\hspace{0.4cm} \psi^i_n =  \dfrac{\sqrt{dP_{01}}}{\sin\theta}\;,
\end{align}
where $P_{kl}$ represents the probability of the projective measurement 
on the basis $|kl\rangle$, i.e., $P_{kl} = |\langle kl |\eta\rangle|^2$ 
of the outcome state $|\eta\rangle$.

To figure out the quality of the reconstructions 
we calculate the trace distance between 
the true state and the reconstructed state as 
$D(\rho^{\rm true},\rho^{\rm rec})=Tr(|\rho^{\rm true}-\rho^{\rm rec}|)/2$ \cite{Fuchs45}.
Evidently, ReDSM significantly reduces the number of projective measurements. 
Therefore, we predict that ReDSM will give more efficient results 
than the  usual DSM \cite{Vallone116,Maccone89}. 
To illustrate, we compare the trace distance $D(\rho^{\rm true},\rho^{\rm rec})$ 
obtained from ReDSM with those ones obtained from the usual DSM and 
the mutually unbiased bases 
quantum state tomography (MUB QST)
through the Monte Carlo method introduced in 
Ref. \cite{Maccone89} (see Appendix \ref{appB}). 
The statistical errors induced from the numerical simulation 
can be reduced by considering a finite number N$_{\rm c}$ of copies of the system.  

\begin{figure} 
\centering
\includegraphics[width=8.6cm]{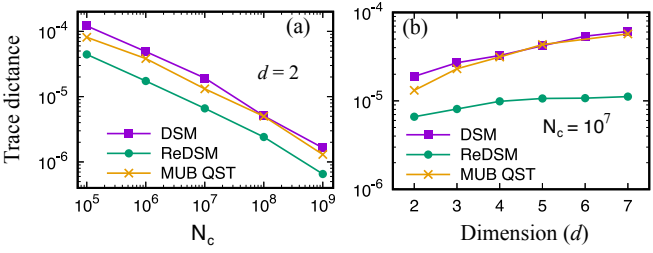}
\caption{
(Color online.) Comparison between the usual DSM 
({\color{Fuchsia}{\scriptsize $\blacksquare$}}),
ReDSM ({\color{ForestGreen} {\large$\bullet$}}), and
MUB QST ({\color{Orange} $\bm{\times$}}).
(a) The trace distances are plotted as functions of 
the total number of copies ${\rm N}_{\rm c}$ for $d = 2$.
(b) The trace distances are plotted as functions of 
the dimension $d$ for fixed ${\rm N}_{\rm c} = 10^7$.
In both cases (a) and (b), the measurement strength 
$\theta= \pi/2$ for the usual DSM and ReDSM, 
while  MUB QST does not depend on $\theta$.
}
\label{fig2}
\end{figure}

Fig. \ref{fig2}(a) shows the trace distances $D(\rho^{\rm true},\rho^{\rm rec})$ 
as functions of a finite number N$_{\rm c}$ 
of copies of the system for the usual DSM,
ReDSM,
and MUB QST. 
We also fixed $\theta= 0.5\pi$ (strong measurement) for the usual DSM
and ReDSM, while MUB QST does not depend on $\theta$. 
We also fixed $d = 2$ in Fig. \ref{fig2}(a).   
Generally, the trace distance decrease with 
the increase of the numbers of copies of the systems N$_{\rm c}$.
We first reproduce that the usual DSM have the same accuracy 
as MUB QST and there are no bias errors as shown in Ref. \cite{Zou91}.
More importantly, we also show that the trace distances obtained from ReDSM 
are more precise than that results obtained from the usual DSM and MUB QST. 
It can be seen that, at the same number of N$_{\rm c}$ of copies, 
the trace distance in ReDSM are smaller than those ones from the usual DSM and MUB QST, 
which means more accuracy. Especially, the bias errors can be eliminated in these cases.
We next show the trace distances $D(\rho^{\rm true},\rho^{\rm rec})$ 
as functions of the dimension $d$ for N$_{\rm c} = 10^7$ in Fig. \ref{fig2}(b). 
Again, the results show that ReDSM is more precise than the usual DSM and MUB QST. 
Furthermore, in cases the usual DSM, ReDSM, and MUB QST, 
the trace distances increase when $d$ increases.

\subsection{For mixed states}
We will next apply our proposal to determine the density matrix $\rho$ directly. 
We consider the mapping \cite{Ho97,Ho}
\begin{align}{\label{roh_tilde}}
\rho\to\widetilde\rho = |0\rangle\langle 0|_e\otimes {\rm Re}(\rho) 
+  |1\rangle\langle 1|_e\otimes {\rm Im}(\rho)\;,
\end{align}
where we have defined 
\begin{align}\label{re_rho}
{\rm Re}(\rho) = 
 \begin{pmatrix}
{\rm Re}(\rho_{0,0}) & {\rm Re}(\rho_{0,1}) & \cdots & {\rm Re}(\rho_{0,d-1}) \\
 {\rm Re}(\rho_{1,0}) & {\rm Re}(\rho_{1,1}) & \cdots & {\rm Re}(\rho_{1,d-1}) \\
  \vdots  & \vdots  & \ddots & \vdots  \\
  {\rm Re}(\rho_{d-1,0}) &{\rm Re}(\rho_{d-1,1}) & \cdots & {\rm Re}(\rho_{d-1,d-1}) 
 \end{pmatrix}\;,
 \end{align}
and similarly for ${\rm Im}(\rho)$. 
We note that in this case, the extra qubit is in the first position 
and the system is in the second position. 
The total state is given as 
$\rho^{\rm in} = \widetilde\rho\otimes|0\rangle\langle 0|_p$. 
The procedure is performed the same as the pure-state case 
with the postselection onto the conjugate basis as 
$|c_j\rangle = 1/\sqrt{d}\sum_{n=0}^{d-1}\omega^{nj}|j\rangle$, 
where $\omega \equiv e^{2\pi i /d}$.
After the postselection, the entangled final state is the two-qubit 
$\rho^{\rm out}$ 
which contains the information of the density matrix, where

\begin{align}{\label{real_out}}
\rho^{\rm out}_{3,0}(n,j) = \dfrac{-i\sin\theta}{d}
\Biggl[\sum_m\rho^r_{m,n}\omega^{(m-n)j}-\epsilon\rho^r_{n,n}\Bigl]
\end{align}
for the real part and 
\begin{align}{\label{ima_out}}
\rho^{\rm out}_{1,2}(n,j) = \dfrac{-i\sin\theta}{d}
\Biggl[\sum_m\rho^i_{m,n}\omega^{(m-n)j}-\epsilon\rho^i_{n,n}\Bigl]
\end{align}
for the imaginary part. Here $\rho^r_{n,n}$ and $\rho^i_{n,n}$ are given in the relations
\begin{align}{\label{nn_out}}
\rho^{\rm out}_{3,3}(n) = \dfrac{\sin^2\theta}{d}\rho^r_{n,n},   \text{and } \rho^{\rm out}_{1,1}(n) = \dfrac{\sin^2\theta}{d}\rho^i_{n,n}\;.
\end{align}
%
Using the Fourier transform, we obtain the desired reconstructed density matrix as
%
\begin{align}
\rho^r_{n,m} &= \dfrac{1}{\sin\theta}\Bigl[d\tan\frac{\theta}{2}\delta_{nm}\rho^{\rm out}_{3,3}(n)+i\sum_j\rho^{\rm out}_{3,0}(n,j)\omega^{(n-m)j}\Bigl],{\label{real_rho_re}}\\
\rho^i_{n,m} &= \dfrac{1}{\sin\theta}\Bigl[d\tan\frac{\theta}{2}\delta_{nm}\rho^{\rm out}_{1,1}(n)+i\sum_j\rho^{\rm out}_{1,2}(n,j)\omega^{(n-m)j}\Bigl].{\label{im_rho_re}}
\end{align}
%
We note that in the mixed-state case, 
we lose the ``tracking information" into the non-diagonal parts of the outcome matrix 
$\rho^{\rm out}$, i.e., $\rho^{\rm out}_{1,2}$ and $\rho^{\rm out}_{3,0}$, 
[which cannot be measured simply by using single projective operations.] 
Therefore, the accuracy can be improved in other ways, 
such as by using various measured bases \cite{Adam105}.
In this subsection, to evaluate the desired outcome states 
$\rho^{\rm out}_{3,0}, \rho^{\rm out}_{3,3}, \rho^{\rm out}_{1,2}$ 
and $\rho^{\rm out}_{1,1}$ we measure them in the two kinds of bases. 
Practically, we use a set of projectors from a pairwise combination 
of eigenstates of the Pauli operators, 
which is known as standard separable state basis (SSB ReDSM) \cite{Nielsen}. 
The desired set of projectors is shown in the left column of Table \ref{tab:title}. 
For comparison, the second basis, which is known as the Bell and Bell-like states basis (BBB ReDSM) 
as we show in the right column of Table \ref{tab:title}, is also used. 
Here, we 
emphasize that these two kinds of bases are applied to ReDSM only.
\begin{table}
\footnotesize
\centering
\newcolumntype{K}[1]{>{\centering\arraybackslash}p{#1}}
\caption {The measurement bases of SSB ReDSM and BBB ReDSM that are used in our scheme. Therein, $|D\rangle = (|0\rangle+|1\rangle)/\sqrt{2}, |A\rangle = (|0\rangle-|1\rangle)/\sqrt{2}, |L\rangle = (|0\rangle+i|1\rangle)/\sqrt{2},|R\rangle = (|0\rangle-|1\rangle)/\sqrt{2}$. } \label{tab:title}
\setlength{\tabcolsep}{6pt}
\begin{tabular}{K{3.0cm} K{4.5cm}}
  \hline\hline
  SSB ReDSM & BBB ReDSM	\\
  \hline
  $|DD\rangle, |DA\rangle, |AD\rangle, |AA\rangle$ & $(|00\rangle+|11\rangle)/\sqrt{2}, (|00\rangle-|11\rangle)/\sqrt{2}$ \\
  $|DL\rangle, |DR\rangle, |AL\rangle, |AR\rangle$ & $ (|00\rangle+i|11\rangle)/\sqrt{2}, (|00\rangle-i|11\rangle)/\sqrt{2}$ \\
  $|LD\rangle, |LA\rangle, |RD\rangle, |RA\rangle$ & $(|01\rangle+|10\rangle)/\sqrt{2},(|01\rangle-|10\rangle)/\sqrt{2}$ \\
  $|LL\rangle, |LR\rangle, |RL\rangle, |RR\rangle$ &  $(|01\rangle+i|10\rangle)/\sqrt{2}, (|01\rangle-i|10\rangle)/\sqrt{2}$\\
  $|00\rangle, |01\rangle$ & $|00\rangle, |01\rangle$ \\
  \hline \hline 
\end{tabular}
\end{table}

\begin{figure} [t]
\centering
\includegraphics[width=8.6cm]{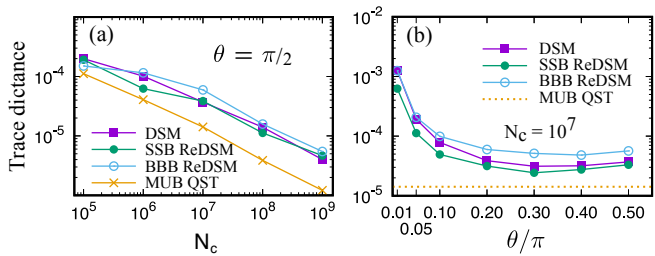} 
\caption{
(Color online.) Comparison between the usual DSM 
({\color{Fuchsia}{\scriptsize $\blacksquare$}}),
ReDSM [including SSB ReDSM ({\color{ForestGreen} {\large$\bullet$}})
and BBB ReDSM ({\color{SkyBlue} {\large$\circ$}})], and 
MUB QST ({\color{Orange} $\bm{\times$}}). 
(a) The trace distances are plotted as functions of N$_{\rm c}$, where $\theta = \pi/2$.
(b) The trace distances are plotted as functions of the measurement strength $\theta$, 
where ${\rm N}_{\rm c} = 10^7$. 
We note that  MUB QST does not depend on $\theta$ 
and is plotted as a dotted line in (b).
In both cases (a) and (b), we fixed $d=2$.
}
\label{fig3}
\end{figure}

Let us first examine the obtained results of the trace distances for fourth cases, 
namely DSM, SSB ReDSM, BBB ReDSM, and MUB QST 
as functions of the number of copies ${\rm N}_{\rm c}$. 
The results are shown in Fig. \ref{fig3}(a) for $\theta = \pi/2$ and $d=2$. 
It can be seen that SSB ReDSM and BBB ReDSM have the same accuracy as the usual DSM.
However, the efficiencies of SSB ReDSM and BBB ReDSM are less than that of MUB QST.
The rebits method does not improve the accuracy in the case of mixed states.

We next consider the coupling strength dependence of the trace distances in  
the usual DSM, SSB ReDSM, and BBB ReDSM in Fig. \ref{fig3}(b). 
We fixed ${\rm N}_{\rm c}=10^7$ and $d=2$.
The results show that with the increasing the coupling strength $\theta$,
all the trace distances decrease steadily. 
The results are in agreement with Vallone's reports \cite{Vallone116} 
saying that strong measurements give better results.
Of course, in MUB QST, there is no interaction with the pointer. 
Therefore, the trace distance of MUB QST is unchanged under the variation of $\theta$
and is showed as the dotted line in Fig. \ref{fig3}(b), where the result is extracted from
Fig. \ref{fig3}(a) at ${\rm N}_{\rm c}=10^7$.

\begin{figure} [t]
\centering
\includegraphics[width=8.6cm]{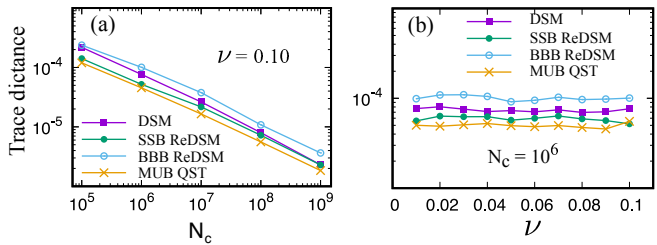} 
\caption{
(Color online.)  Comparison between the usual DSM 
({\color{Fuchsia}{\scriptsize $\blacksquare$}}),
ReDSM [including SSB ReDSM ({\color{ForestGreen} {\large$\bullet$}})
and BBB ReDSM ({\color{SkyBlue} {\large$\circ$}})], and 
MUB QST ({\color{Orange} $\bm{\times$}}). 
(a) The trace distances are plotted as functions of N$_{\rm c}$, where $\nu = 0.10$.
The apparent coincidence of SSB ReDSM and DSM at high N$_{\rm c}$ due to numerical errors.
(b) The trace distances are plotted as functions of $\nu$, where ${\rm N}_{\rm c} = 10^6$.
In both cases (a) and (b), $\theta = \pi/2$ and $d=2$.
}
\label{fig4}
\end{figure} 

\subsection{For nearly-pure states}
Remarkably, the results in Fig. \ref{fig3} suggest that 
SSB ReDSM appears to have the lowest trace distance
in comparison to the usual DSM and BBB ReDSM. 
For more concreteness, we consider the case of nearly-pure states, 
whereby, $\rho = (1-\nu)|\psi\rangle\langle\psi| +\nu \bm{I}/d$, 
where $|\psi\rangle$ is a pure state, $\nu$ small noise.
In practice, nearly-pure states usually appear 
due to some noises in the preparation stage \cite{Martinez119,Zhang2018}. 
The comparison between the usual DSM, SSB ReDSM, BBB ReDSM, and MUB QST 
is given in Fig. \ref{fig4}(a) for $\nu = 0.10$.
We first observe that among the DSM variations 
(the usual DSM, SSB ReDSM, and BBB ReDSM,) 
SSB ReDSM provides higher accuracy. 
It appears to coincide to the usual DSM for larger ${\rm N}_{\rm c}$
due to numerical errors.
Furthermore, the accuracy of  SSB ReDSM is asymptotic to MUB QST.
It suggests that the rebit method can be useful for reconstruction nearly-pure states 
in the scene that QST is difficult to realize. 

In Fig. \ref{fig4}(b), we consider a wide range of the noise $\nu$ from 0.01 to 0.10 
for a fixed ${\rm N}_{\rm c} = 10^6$.
The results show that SSB ReDSM offers the best accuracy 
in comparison to the others DSM variations.
SSB ReDSM asymptotic to MUB QST is also observed. 
Obviously, the rebit method can assist the robustness to noises.

\begin{figure} [t]
\centering
\includegraphics[width=8.6cm]{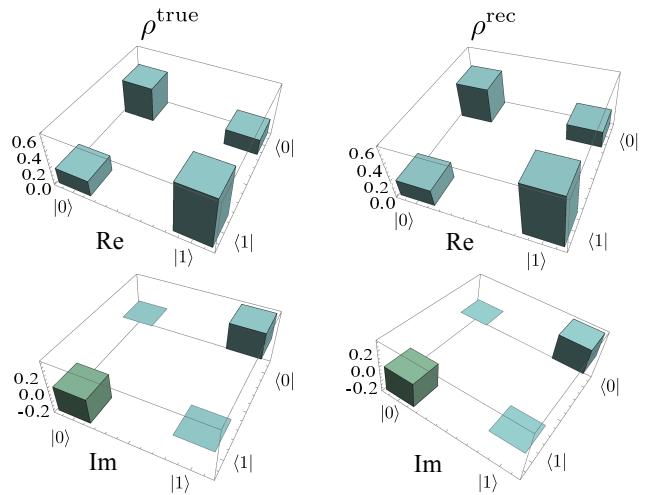} 
\caption{
(Color online.) A random-generated mixed state (left column) 
and the reconstructed state (right column).
Re and Im are ``real" and ``imaginary," respectively.
}
\label{fig5}
\end{figure}

Finally, we also emphasize that the direct state measurements (including the rebit method)
might not provide physical reconstructed states, i.e.,
the reconstructed mixed state might not have positive eigenvalues.
To resolve this issue, one might need to perform additional techniques like a maximum-likelihood estimation.
Hereafter, we check that (by using Monte-Carlo simulation) the rebit method gives a physical density matrix, 
i.e., have positive eigenvalues.
As an example, we consider a random-generated mixed state as shown in the left column Fig. \ref{fig5}, where
\begin{align}\label{rhot}
\rho^{\rm true} =
\begin{pmatrix}
0.40693 & 0.18711 +0.32119i\\
0.18711 -0.32119i &0.59307
\end{pmatrix}
\end{align}
This state is normalized, i.e., $Tr(\rho^{\rm true}) =1$, 
and Hermitian, i.e., $\rho^{\rm true}=[\rho^{\rm true}]^\dagger$.
The eigenvalues are 0.88319 and 0.11681, positive and sum unity. 
The corresponding reconstructed state is given in the right column of Fig. \ref{fig5}, where
\begin{align}\label{rhore}
\rho^{\rm rec} =
\begin{pmatrix}
0.40694 & 0.18711 +0.32121i\\
0.18711-0.32121i &0.59306
\end{pmatrix},
\end{align}
where the SSB ReDSM method was used to reproduce this state. 
We show that the reconstructed state is also a physical state, 
which guarantees that its eigenvalues, 0.88321 and 0.11679, are positive.

\sloppy
\section{Conclusions}{\label{sec_iv}
We have proposed a method for improving the accuracy in direct state measurements. 
By adding an extra qubit, 
we can decompose and keep tracking the real and imaginary parts of the initial unknown state. 
After the interaction and postselection the system into the conjugate bases, 
the extra qubit and pointer qubit outcomes can be measured 
and taken out the reconstructed state. 
For pure states, on one hand, 
it is easy to perform projective measurements on the output state. 
For mixed states, on the other hand, 
both cases the standard separable state basis (SSB ReDSM) 
and the Bell and Bell-like states basis (BBB ReDSM) have been used and compared.   

Our numerical results are concentrated on showing 
the influences of the trace distance between 
the true state and the reconstructed state by the finite number of copies of the system. 
For pure states, the trace distances obtained from the rebits method 
are more precise than those ones obtained from the previous methods, (i.e.,
standard direct state measurements (DSM) and mutually unbiased bases quantum state tomography (MUB QST),) 
because the number of projective measurements is reduced in comparison to the previous methods.
For mixed states, the rebits method with both SSB ReDSM and BBB ReDSM 
have the same accuracy as the standard DSM,
while they are less precise than MUB QST.
However, for nearly-pure states, BBB ReDSM of the rebit method offers a better accuracy, 
which is asymptotic to  MUB QST.
Despite its less efficient in comparison to QST in mixed states, 
the rebit method is much easier to be manipulated. 
Thus, it may offer benefits to be applied.
We believe that our method gives a better solution for 
quantum state tomography and provides a reliable tool for 
universal quantum computing that uses only real amplitudes.  

\section{Acknowledgments}
We would like to thank Yasushi Kondo of Kindai University for useful discussion. 
This work was supported by CREST(JPMJCR1774), JST.

 \appendix
\section{The interaction implementation}\label{appA}
\setcounter{equation}{0}
\renewcommand{\theequation}{A\arabic{equation}}

In this Appendix, we will show how to implement the interaction 
$\bm{\widetilde U}_n(\theta)$ in Eq. \eqref{operator}. 
For simplicity, let us consider the qubit system only ($d = 2$). 
Concretely, there are three qubits in the total system: 
the system qubit, the extra qubit, and the pointer qubit. 
The interaction \eqref{operator}  yields
\begin{align}{\label{operator_App}}
\bm{\widetilde U}_n(\theta)&={\rm exp}\Bigl[{i\frac{\theta}{2}(I_s\pm\sigma^z_s)\otimes\sigma^y_e\otimes\sigma^y_p}\Bigr]\;,
\end{align}
where $\pm$ corresponds to $|n\rangle = |0\rangle$ and $|1\rangle$, respectively. We expand  
\begin{align}{\label{operator_App_Trot}}
\bm{\widetilde U}_n(\theta)&={\rm exp}\Bigl({i\frac{\theta}{2}I_s\otimes\sigma^y_e\otimes\sigma^y_p}\Bigr){\rm exp}\Bigl({\pm i\frac{\theta}{2}\sigma^z_s\otimes\sigma^y_e\otimes\sigma^y_p}\Bigr).
\end{align}
The first term is a usual two-body interaction, therefore it can be implemented easily. 
To implement the second term, we use a sequence of nonlocal entangling 
M\o lmer-S\o rensen gates and local single-qubit rotations, where 
\begin{align}{\label{MS}}
\bm{U}_{\rm MS}(\phi,\varphi)={\rm exp}[-i\frac{\phi}{4}(\cos\varphi \bm{S}_x+\sin\varphi \bm{S}_y)^2], 
\end{align}
is the M\o lmer-S\o rensen gate  \cite{Muller13, Molmer82}, 
where $\bm{S}_{x,y}=\sum_{i=1}^N\bm{\sigma}^i_{x,y}$, $\phi$ 
and $\varphi$ are two angle parameters, and $N$ is the number of local qubits. 
%
%
We first (i) apply the M\o lmer-S\o rensen gate to all: 
the system qubits, the extra qubit, and the pointer qubit, 
then (ii) apply a  local single-qubit rotation onto the pointer qubit, 
and finally (iii) the M\o lmer-S\o rensen gate is applied again (see Fig. \ref{fig1}.) 
This three-step implementation can be seen from Refs. \cite{Casanova108, Ho97}.  
Following Ref. \cite{Casanova108}, we have
\begin{align}{\label{operator_App_2}}
e^{{\pm i\frac{\theta}{2}\sigma^z_s\otimes\sigma^y_e\otimes\sigma^y_p}} = \bm{U}_{\rm MS}(-\dfrac{\pi}{2},\dfrac{\pi}{2})e^{\pm i\frac{\theta}{2}\sigma^z_p}\bm{U}_{\rm MS}^\dagger(-\dfrac{\pi}{2},\dfrac{\pi}{2})\;.
\end{align}
We emphasize that these gates can be simulated by various physical platforms, 
such as ion traps, quantum photonics, superconducting circuits, and others.

\section{The Monte Carlo simulation scheme}\label{appB}
\setcounter{equation}{0}
\renewcommand{\theequation}{B\arabic{equation}}
We follow the Monte Carlo simulation scheme described in Ref. \cite{Maccone89}.

\subsection{For pure states}

\textbf{Step 1.} Prepare a quantum state $|\psi\rangle$ for a fixed of $d$-dimension. 
In this work, we generate a quantum state randomly. 
We also choose the finite number N$_{\rm c}$ of copies and the coupling strength $\theta$.

\textbf{Step 2.} Evaluate the postselection probability 
$P_0 = |\langle c_0|\psi\rangle|^2 = \frac{1}{d}\sum_{n=0}^{d-1}(\psi^r_n)^2+(\psi^i_n)^2$.

\textbf{Step 3.} For $n$ from 0 to $d-1$ do:
\begin{itemize}
  \item For $i$ from 0 to N$_{\rm c}P_0/d$ do: 
  (Notably, for each fixed N$_{\rm c}$ of copies, we just run N$_{\rm c}P_0/d$ 
  of copies because the postselection onto $|c_0\rangle$ discards all other cases.)
  \begin{itemize}
     \item Generate a random number distributed according to 
     the probability distribution for the extra qubit  using the cumulative method 
     and bisection method \cite{Maccone89}.
     \item Collect the results of $P_{01}(n)$ and $P_{11}(n)$ from the simulation. 
  \end{itemize}
  \item Average the simulated results to get the estimated state (for each $n$.)
\end{itemize}

\textbf{Step 4.} Normalize the estimated state and evaluate the trace distance. 


 \subsection{For mixed states}

\textbf{Step 1.} Prepare a random quantum mixed state 
$\rho$ for a fixed of $d$-dimension. Notable that the state is normalized, 
i.e., $Tr(\rho) =1$, and Hermitian, i.e., $\rho=\rho^\dagger$.

\textbf{Step 2.} For $n$ from 0 to $d-1$ do:
\begin{itemize}
 \item Evaluate the postselection probability $P_j = \langle c_j|\rho|c_j\rangle$. 
 Then, using the cumulative method to generate the corresponding $j$, $j \in [0, d-1]$. 
 We emphasize that in this case, we do not discard the results 
 because each outcome of the system state will have a 
 corresponding postselected state $|c_j\rangle$.
 \item For $i$ from 0 to N$_{\rm c}/d$ do: 
 \begin{itemize}
  \item Generate a random number according to the cumulative method 
  and bisection method to evaluate the outcomes $\rho^{\rm out}_{p,q} (n,j)$
   as given in Eqs. (\ref{real_out}-\ref{nn_out}). Here, $p,q = \{(3,0); (1,2); (3,3); (1,1).\}$
  \item Collect the simulated results.
   \end{itemize}
   \item Average the simulated results for each $j$ and $n$. 
   Then, using the Fourier transform to evaluate the estimated state as given in Eqs. (\ref{real_rho_re}, \ref{im_rho_re}).
\end{itemize}

\textbf{Step 3.} Normalize the estimated state and evaluate the trace distance. 

Notable that to calculate the statistical errors,
we run the scheme $M$ times with N$_{\rm c}/M$ number of copies for each time.
Of course, at each time, a random state will be generated. Here, we choose $M = 100$.

\end{document}